\documentclass[prb,twocolumn,showpacs,superscriptaddress,amsmath,floatfix]{revtex4}
\usepackage{graphicx}
\usepackage{amssymb}
\usepackage{citesort}
\usepackage{dcolumn}

\begin{document}
\title{Absence of long-range superconducting correlations in
 the frustrated $\frac{1}{2}$-filled  band Hubbard model}
\author{S. Dayal}
\affiliation{Department of Physics and Astronomy and HPC$^2$ Center for Computational 
Sciences, Mississippi State University, Mississippi State, MS, 39762}
\author{R.T. Clay}
\affiliation{Department of Physics and Astronomy and HPC$^2$ Center for Computational 
Sciences, Mississippi State University, Mississippi State, MS, 39762}
\author{S. Mazumdar}
\affiliation{Department of Physics, University of Arizona, Tucson, AZ, 85721}
\begin{abstract}
We present many-body calculations of superconducting pair-pair
correlations in the ground state of the half-filled band Hubbard model
on large anisotropic triangular lattices. Our calculations cover
nearly the complete range of anisotropies between the square and
isotropic triangular lattice limits.  We find that the superconducting
pair-pair correlations decrease monotonically with increasing onsite
Hubbard interaction $U$ for inter-pair distances greater than nearest
neighbor.  For the large lattices of interest here the distance
dependence of the correlations approaches that for noninteracting
electrons.  Both these results are consistent with the absence of
superconductivity in this model in the thermodynamic limit.  We
conclude that the effective $\frac{1}{2}$-filled band Hubbard model,
suggested by many authors to be appropriate for the
$\kappa$-(BEDT-TTF)-based organic charge-transfer solids, does not
explain the superconducting transition in these materials.
\end{abstract}
\pacs{71.10.Fd,71.30.+h,74.20.Mn}
\maketitle

\section{Introduction}
\label{intro}
The two dimensional (2D) Hubbard model has been extensively
investigated because at $\frac{1}{2}$-filling it can successfully
describe the antiferromagnetic (AFM) phases found in many
strongly-correlated materials. Since AFM phases often occur in
materials displaying unconventional superconductivity (SC), such as
the high-$T_c$ cuprates and the organic $\kappa$-(BEDT-TTF)$_2$X
(hereafter $\kappa$-(ET)$_2$X) charge transfer solid (CTS)
superconductors, it has frequently been suggested that some small
modification of the model can yield a superconducting state where the
residual AFM fluctuations mediate an attractive pairing interaction.
In the case of the cuprates, this modification involves a change in
the carrier concentration (``doping''); the doped 2D Hubbard model has
been intensively investigated with numerous analytic and numerical
methods, but whether or not SC occurs within this model is still
controversial.

\begin{figure}[tb]
\centerline{\resizebox{2.5in}{!}{\includegraphics{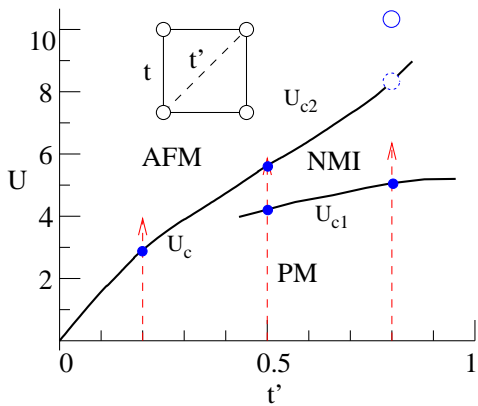}}}
\caption{(color online) Lattice structure (see inset) and
  the ground state phase diagram of the
  $\frac{1}{2}$-filled band Hubbard model on the anisotropic triangular
  lattice for $0\leq t^\prime < 1$. Phases labeled are paramagnetic
  metallic (PM), N\'eel antiferromagnetic (AFM), and non-magnetic
  insulator (NMI).  $d_{x^2-y^2}$ superconductivity has been suggested
  to occur near the boundary of the AFM and PM phases (see text).
Filled circles are finite-size scaled values for phase boundaries 
from PIRG calculations
(see Section \ref{results}). 
The precise value of $U_{c2}$ at $t^\prime=0.8$ is known with less certainty.
At $t^\prime=0.8$ the solid circle is
an exact upper bound from the 4$\times$4 lattice, while the
dashed circle  shows the expected finite-size scaled value.
The phase boundary lines linking the  points are only schematic guides to the eye.
Dashed vertical 
lines indicate the parameter regions
we 
investigate for superconductivity.}
\label{phasediag}
\end{figure}

The AFM state in the 2D Hubbard model can also be destroyed at fixed
carrier concentration by the introduction of lattice frustration. The
model on the anisotropic triangular lattice (see Fig.~\ref{phasediag})
has been used to describe the $\kappa$-(ET)$_2$X family of CTS, where
SC occurs at fixed carrier density under application of moderate
pressure.  The ET layers here consist of strongly dimerized
anisotropic triangular lattices, with the intradimer hopping integrals
much larger than the interdimer ones. Each (ET)$_2^+$ dimer contains
one hole carrier on the average. This has been used to justify
replacing each dimer unit cell with a single site, and the underlying
$\frac{1}{4}$-filled cation band with an {\it effective}
$\frac{1}{2}$-filled band \cite{Kanoda11a}.

We investigate ground state superconducting pair-pair correlations within the Hamiltonian,
\begin{eqnarray}
H & = & -t\sum_{\langle ij\rangle,\sigma} (c^\dagger_{i,\sigma}c_{j,\sigma}+H.c.)  
-t^\prime\sum_{[kl],\sigma} (c^\dagger_{k,\sigma}c_{l,\sigma}+H.c.) \nonumber \\
&+&U\sum_i n_{i,\uparrow}n_{i,\downarrow}.
\label{ham}
\end{eqnarray}
In Eq.~\ref{ham}, $c^\dagger_{i,\sigma}$ creates an electron of spin
$\sigma$ on site $i$ and
$n_{i,\sigma}=c^\dagger_{i,\sigma}c_{i,\sigma}$. $U$ is the on-site
Hubbard interaction.  We consider a square lattice with hopping
integral $t$ along $x$ and $y$-directions and frustrating hopping
$t^\prime$ along the $x+y$-direction (see Fig.~\ref{phasediag}).  The
limits $t^\prime/t=0$ and 1 correspond to the square and the isotropic
triangular lattices, respectively.  All quantities with dimensions of
energy will be expressed hereafter in units of $t$. We consider only
the $\frac{1}{2}$-filled band corresponding to an electron density per
site $\rho=1$.

The non-superconducting phases of this model, shown schematically in
Fig.~\ref{phasediag}, are relatively well established.  As $t^\prime$
is increased in strength, frustration destroys the {\bf
  q}=($\pi$,$\pi$) AFM ground state, replacing it with either a
paramagnetic metallic (PM) state or a non-magnetic insulator (NMI)
state \cite{Kashima01a,Morita02a}.  Numerical calculations on this
model and the related model with two diagonal $t^\prime$ bonds in each
plaquette have confirmed the presence of the PM, AFM, and NMI phases
\cite{Kashima01a,Morita02a,Watanabe03b,Parcollet04a,Mizusaki06a,Kyung06a,Sahebsara06a,Koretsune07a,Clay08a,Lee08b,Sahebsara08a,Watanabe08a,Ohashi08a,Tocchio09a,Yoshioka09a,Guertler09a,Liebsch09a,Yu10a}.
The NMI phase has been suggested as a candidate state \cite{Morita02a}
that explains the apparent quantum spin liquid (QSL) behavior seen in
the strongly frustrated $\kappa$-(ET)$_2$Cu$_2$(CN)$_3$
\cite{Kanoda11a}. As the NMI phase has already been extensively
investigated
\cite{Morita02a,Watanabe03b,Mizusaki06a,Koretsune07a,Tocchio09a,Yoshioka09a},
in the present work we will not consider the properties of this phase
any further, but will rather focus on the possibility of SC within the
model.

Numerous mean-field theories have suggested that unconventional SC
occurs adjacent to AFM-PM phase boundaries
\cite{Schmalian98a,Kino98a,Kondo98a,Vojta99a,Baskaran03a,Liu05a,Kyung06a,
  Sahebsara06a,Nevidomskyy08a}.  Similar superconducting states have
been suggested for the closely related Hubbard-Heisenberg model on the
same lattice \cite{Gan05a,Powell05a,Gan06a,Powell07a,Rau11a}.  Because
of the proximity of ($\pi$,$\pi$) AFM order, the suggested symmetry of
the SC order parameter is $d_{x^2-y^2}$.  For $t^\prime\sim1$ the
magnetic ordering {\bf q} shifts to
($\frac{2\pi}{3}$,$\frac{2\pi}{3}$) corresponding to the 120$^\circ$
ordering found in the triangular lattice antiferromagnetic Heisenberg
model, and consequently superconducting other order parameter
symmetries have been suggested \cite{Powell07a}.  The estimated value
of $t^\prime$ for the $\kappa$-ET materials is however smaller than 1
(see below) \cite{Kandpal09a,Nakamura09a}, and also no evidence for
120$^\circ$ AFM order is found experimentally within the $\kappa$-ET
family \cite{Kanoda11a}.

Superconducting pair-pair correlations calculated with numerical
methods going beyond mean field theory provide a more accurate
assessment of the presence of SC, provided finite-size effects can be
adequately controlled.  Two criteria must be satisfied to confirm SC
within the model: (i) the superconducting pair-pair correlations must
be enhanced over the $U=0$ values over at least a range of $U$, and
(ii) the pair-pair correlations must extrapolate to a finite value at
long inter-pair distances.  We have previously calculated pair-pair
correlations for the 4$\times$4 lattice using exact diagonalization
\cite{Clay08a}.  No enhancement of the pair-pair correlations by $U$
was found in these calculations, except for a trivial short-distance
enhancement \cite{Clay08a} (see also below).  Our present work allows
more careful analyses of the distance dependence of the pair-pair
correlations, as well as the U-dependence of the longer-range
components of these correlations, that were not possible within the
earlier small cluster calculation.

Pair-pair correlations for lattices comparable to those in the present
work have also been calculated using variational quantum Monte Carlo
(VMC) methods
\cite{Watanabe06a,Yokoyama06a,Watanabe08a,Tocchio09a,Guertler09a}.
VMC results however depend to a great degree on the choice of the
variational wavefunction, and there is considerable differences of
opinion within the existing VMC literature.  Clearly, studies of
pair-pair correlations on large lattices, using many-body methods that
do not depend on an a priori choice of the wavefunction are desirable.
A candidate method for calculations of strongly-correlated systems is
the recently developed Path Integral Renormalization Group (PIRG)
method \cite{Imada00a,Kashima01b,Mizusaki04a,Imada06a}.  Like VMC,
PIRG is also variational and does not suffer from a fermion sign
problem as do standard quantum Monte Carlo methods. Unlike VMC methods
however, instead of an assumed functional form of the wavefunction,
PIRG uses an {\it unconstrained} sum of Slater determinants that is
optimized using a renormalization procedure
\cite{Imada00a,Kashima01b,Mizusaki04a,Imada06a}.  The NMI phase within
Eq.~\ref{ham} was first identified using PIRG
\cite{Kashima01a,Morita02a}.  Previous PIRG calculations
\cite{Kashima01a,Morita02a,Mizusaki06a,Watanabe03b,Yoshioka09a}
investigated the metal-insulator transition, AFM ordering, and
properties of the NMI phase in detail, but did not discuss
superconducting pair-pair correlations. Here we revisit the model with
PIRG and calculate pair-pair correlations as a function of $t^\prime$
and $U$.  As explained in Section \ref{pirg}, we use the most accurate
version of the PIRG ground-state method, Quantum Projection-PIRG
(QP-PIRG), which combines symmetries with the renormalization
procedure \cite{Mizusaki04a}.  As explained in Section \ref{pirg}, we
also performed an ``annealing'' procedure to help prevent the method
from converging to local minima.

While early tight-binding bandstructures calculated using the extended
H\"uckel method found some $\kappa$-ET superconductors to have nearly
isotropic triangular lattices with $t^\prime\approx 1$, recent
ab-initio methods have determined that $t^\prime$ in the experimental
systems lie within the range $0.4 \alt t^\prime \alt 0.8$
\cite{Kandpal09a,Nakamura09a}. Importantly, in this range of
anisotropy the 120$^\circ$ AFM order is not relevant. Furthermore, the
AFM order is known experimentally to be of the conventional Ne\'el
pattern \cite{Miyagawa95a,Miyagawa02a}. Consequently, we limit our
calculations to $t^\prime\alt 0.8$.  Specifically, we perform our
calculations for three distinct $t^\prime=0.2$, 0.5, and 0.8, as shown
in Fig.~\ref{phasediag}. The two large $t^\prime$ values chosen
bracket the estimated frustration in the $\kappa$-ET superconductors
\cite{Kandpal09a,Nakamura09a}.  We choose a smaller $t^\prime=0.2$ in
addition because it has been suggested in some studies that SC is
present even in the weakly frustrated region of the phase diagram
\cite{Powell07a,Nevidomskyy08a,Guertler09a}.  The phase diagram
(Fig.~\ref{phasediag}) is qualitatively different at $t^\prime=0.2$
because the NMI phase does not occur for $t^\prime<0.5$
\cite{Kashima01a,Morita02a,Mizusaki06a}.  While the estimate for
degree of frustration is remarkably consistent between different
ab-inito methods \cite{Kandpal09a,Nakamura09a}, the estimated value of
$U$ for $\kappa$-(ET)$_2$X is less certain.  We therefore perform our
calculations over a range of $U$ starting from $U=0$.

The organization of the paper is as follows. In Section \ref{oparams},
we introduce definitions of the order parameters we calculate. In
Section \ref{pirg} we describe the PIRG method. Section \ref{results}
presents our data for $t^\prime=0.2$, 0.5, and 0.8, followed by
discussions and conclusion in Section \ref{discussion}.

\section{Order parameters}
\label{oparams}
	
To determine whether SC is present near the metal-insulator (MI)
transition, in addition to superconducting correlations we need order
parameters to distinguish between metallic and insulating phases.  To
locate the MI transition we use two different quantities. The first is
the double occupancy $D=\langle n_{i\uparrow}n_{i\downarrow}\rangle$.
As $U$ increases, a discontinuous decrease in $D$ occurs at the MI
transition \cite{Morita02a}. In addition, we calculate the bond order
$B_{ij}$ between sites $i$ and $j$,
\begin{equation}
B_{ij}=\sum_\sigma\langle c^\dagger_{i,\sigma}c_{j,\sigma}+H.c.\rangle.
\end{equation} 
In the following we have labeled the bond order between sites linked by the $t^\prime$ bond
as $B^\prime$. This particular bond order is
nonzero in the PM phase but tends to zero in the AFM Ne\'el ordered
phase because there electrons on sites connected by $t^\prime$
have parallel spin projections \cite{Clay08a}. 

The operator $\Delta^\dagger_{i,j}$ creates a singlet pair on
lattice sites $i$ and $j$:
\begin{equation}
\Delta^\dagger_{i,j}= \frac{1}{\sqrt{2}}
(c^\dagger_{i,\uparrow}c^\dagger_{j,\downarrow}
- c^\dagger_{i,\downarrow}c^\dagger_{j,\uparrow}).
\end{equation}
The pair-pair correlation function is defined as
\begin{equation}
P_\alpha({\bf r})=\frac{1}{4}\sum_\nu g(\nu)\langle \Delta^\dagger_i
\Delta_{i+\bf{r}(\nu)}\rangle.
\label{pair}
\end{equation}
In Eq.~\ref{pair} the phase factor $g(\nu)$ determines the symmetry of
the superconducting order parameter.  We consider two possible pairing
symmetries, $d_{x^2-y^2}$ pairing ($\alpha=d$ in our nomenclature
below) where $g(\nu)=\{+1,-1,+1,-1\}$ and ${\bf
  r}(\nu)=\{\hat{x},\hat{y},-\hat{x},-\hat{y}\}$, and $d_{xy}$ pairing
($\alpha=xy$) where $g(\nu)=\{+1,-1,+1,-1\}$ and ${\bf
  r}(\nu)=\{\hat{x}+\hat{y},-\hat{x}+\hat{y},
-\hat{x}-\hat{y},\hat{x}-\hat{y}\}$.

In the presence of superconducting long-range order, $P_\alpha({\bf
  r})$ for the ground state in the appropriate pairing channel must
converge to a nonzero value for $|{\bf r}|\rightarrow\infty$.  This is
seen clearly for example in the 2D $-U$ Hubbard model
\cite{Scalettar89b,Moreo92b,Huscroft98a}.  In the thermodynamic limit
the long-distance limit of the pair-pair correlation function,
$P_\alpha(r\rightarrow\infty)$, is proportional\cite{Aimi07a} to the
square of the superconducting order parameter, $\langle
\Delta_\alpha\rangle^2 \propto |P(_\alpha(r\rightarrow\infty)|$.  The
magnitude of $\langle \Delta_\alpha\rangle$ may further be used to set
limits\cite{Aimi07a} on the superconducting condensation energy, gap
amplitude, and $T_c$.

\section{Method}
\label{pirg}

The PIRG method has been previously used for a variety of
strongly-correlated systems including the 2D Hubbard model
\cite{Kashima01b}, $\frac{1}{2}$-filled frustrated 2D Hubbard models
\cite{Kashima01a,Morita02a,Mizusaki04a,Mizusaki06a,Yoshioka09a}, and
the $\frac{1}{2}$-filled Hubbard model on the checkerboard lattice
\cite{Yoshioka08a,Yoshioka08b}.  Details of the method are well
described in these references. Here we discuss details of our PIRG
implementation, and present comparisons with exact results which
demonstrate the accuracy of the method for calculating pair-pair
correlations.

\begin{figure}
\resizebox{3.3in}{!}{\includegraphics{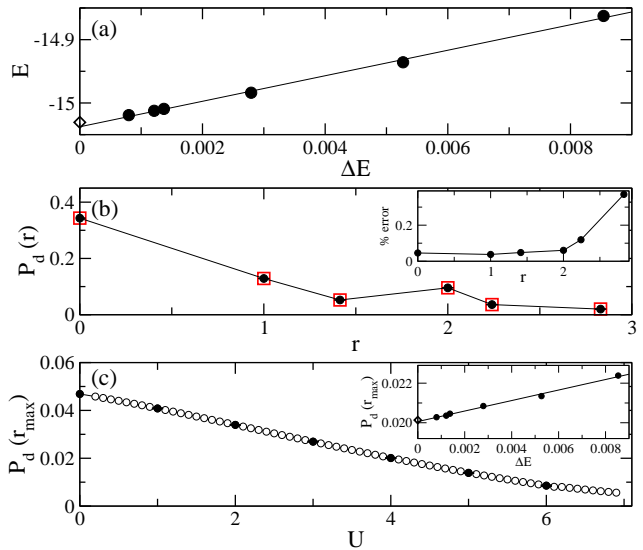}}
\caption{(color online) Comparison of PIRG and exact results for the
  4$\times$4 lattice with $t^\prime=0.5$. (a) Variance extrapolation
  of the ground state energy for $U=4$. The symbol at $\Delta$E=0 is
  the exact energy; the line is a linear fit. (b) $P_d(r)$ versus $r$.
  Open (filled) symbols are exact (PIRG) results. The inset shows
  relative percent error in $P_d(r)$ as a function of $r$.  (c)
  $P_d(r)$ at the largest possible pair spacing on the finite lattice,
  $P_d(r_{\rm max})$, as a function of $U$. Open (filled) symbols are
  exact (PIRG) results.  The inset shows the variance extrapolation of
  $P_d(r_{\rm{max}})$ for $U=4$.}
\label{4x4compare}
\end{figure}

The PIRG method uses a basis of $L$ Slater determinants,
$|\phi_i\rangle$. For $L=1$ this coincides with the Hartree-Fock (HF)
approximation. In practical calculations, maximum $L$'s of a few
hundred are used. The method is initialized with the $L=1$ HF
wavefunction or a similar random starting wavefunction, and the PIRG
renormalization procedure \cite{Kashima01b} is used to minimize the
energy by optimizing through the action of the operator $\exp(-\tau
H)$.  One potential problem with the PIRG renormalization procedure is
that the method can become trapped in a local minimum and not reach
the true ground state \cite{Yoshioka08a}.  Yoshioka {\it et al.}
introduced a technique for avoiding local trapping by introducing
global modifications to the wavefunction \cite{Yoshioka08a}.
Following Ref. \onlinecite{Yoshioka08a}, we also introduced similar
global modifications of the PIRG wavefunctions (``iteration A'' in
Ref. \onlinecite{Yoshioka08a}), which modify determinants in a global
manner by acting on the wavefunction with $\exp[-\tau H]$ defined by a
{\it random} set of Hubbard-Stratonovich variables. In addition, we
also introduced updates to the $|\phi_i\rangle$ based on adding a
random variation to the matrix elements of $[\phi_i]_{j,k}$.  The
amplitude of the variations is decreased systematically in a manner
similar to simulated annealing. We found the addition of these two
global updates to significantly improve the accuracy of the results.

We also incorporated lattice and spin-parity symmetries in the
calculation \cite{Mizusaki04a}. Reference \onlinecite{Mizusaki04a}
introduced two different methods of using symmetry projection: (i)
PIRG-QP, where symmetry projectors are applied to the ground state
wavefunction after it has optimized using PIRG; and (ii) QP-PIRG,
where symmetry projectors are applied at each step of the PIRG
optimization. Here we have used the second more accurate of these two
approaches, QP-PIRG. The lattice symmetries we used included
translation, inversion, and mirror-plane symmetries, a total of 4$N$
symmetries where $N$ is the number of lattice sites.  We also applied
the spin-parity projection operator after the PIRG process. An
advantage of QP-PIRG is that much smaller basis sizes $L$ can be used
\cite{Mizusaki04a}.

Following reference \onlinecite{Kashima01b} we define the energy
variance $\Delta
E=(\langle\hat{H}^2\rangle-{\langle\hat{H}\rangle}^2)/\langle\hat{H}\rangle^2$.
$\Delta E$ is used to correct for the finite basis size $L$.  For each
set of parameters we performed the annealing and A iterations for
successively larger basis sizes $L$. Each correlation function was
then extrapolated to $\Delta E=0$ by performing a linear fit.  The
error bars reported in our results are the standard errors estimated
from the linear fit. The largest $L$ used here for 6$\times$6 and
8$\times$8 lattices was 256.  The smallest $L$ results we used in the
fitting process depended on the value of $U$: for $U\alt2$ we found
that even $L$ as small as 4 fit gave a good linear variance
extrapolation, while for larger $U$ we only used $L\agt 16$ results in
the fit.

In Fig.~\ref{4x4compare}, we compare results from our PIRG code with
exact diagonalization results for the 4$\times$4 lattice
\cite{Clay08a}. Here $L$ of up to 256 were used.
Fig.~\ref{4x4compare}(a) shows the variance extrapolation of the
energy for $t'=0.5$. The extrapolated value for ground state energy is
-15.037$\pm$0.002 compared to the exact ground state of -15.031.  In
Fig.~\ref{4x4compare}(b) we plot the pair-pair correlation $P_d(r)$ as
a function of distance for $U$=4. The inset shows the percent relative
error in $P_d(r)$ as a function of $r$. The maximum relative error is
for $r=r_{\rm{max}}$=2$\sqrt2$ and is smaller than
0.4$\%$. Fig.~\ref{4x4compare}(c) shows the d-wave correlation at the
furthest distance, $P_d(r_{\rm{max}})$ as a function of $U$ for
4$\times$4, $t'=0.5$.  The inset here shows the variance extrapolation
for $P_d(r_{\rm{max}})$ for $U$=4. Again, as in
Fig.~\ref{4x4compare}(a) the extrapolation of the physical quantity is
well within the statistical error.
	
Our PIRG code was further verified against quantum Monte Carlo (QMC)
results for larger lattices in the $t^\prime=0$ limit where QMC does
not suffer from the fermion sign problem at $\frac{1}{2}$-filling.
For the $6\times6$ lattice, the QMC estimate for the ground state
energy \cite{Mizusaki04a} of Eq.~\ref{ham} with $U=4$ and $t^\prime=0$
is $E$= -30.87$\pm$0.05.  Previous QP-PIRG calculations using lattice
translations and spin-parity during the PIRG projection process,
followed by a total-spin $S=0$ projection obtained\cite{Mizusaki04a}
$E$= -30.879.  The extrapolated energy with our choice of 4$N$ lattice
symmetries, spin-parity projection, and a maximum $L$ of 256 was
almost identical, $E$=-30.89$\pm$0.04.

\section{Results}
\label{results}

\begin{figure}
\resizebox{3.3in}{!}{\includegraphics{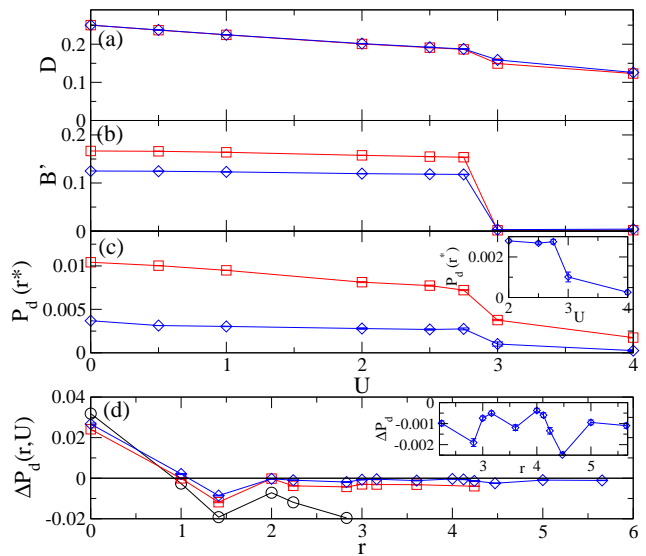}}
\caption{(color online) PIRG results for t'=0.2. Squares and diamonds
  are for 6$\times$6 and 8$\times$8 lattices respectively.  (a) double
  occupancy $D$ (b) $t^\prime$ bond orders (c) long distance
  $d_{x^2-y^2}$ pair-pair correlation function, $P_d(r^\star)$ as a
  function of $U$ (see text).
  The inset shows  
$P_d(r^\star)$ for the 8$\times$8 lattice 
near the PM-AFM boundary.
 (d) Enhancement over non-interacting
  system, $\Delta P_d(r,U)$, for $U=2.75$ as a function of
  distance. Circles here are exact $4\times4$ results.  
  The inset shows 
the long-range part of
$\Delta P_d(r,U)$ 
for the  8$\times$8 lattice.
In all panels lines are only guides to the eye.}
\label{tp0.2-result}
\end{figure}

\subsection{$t'=0.2$}
\label{results-tp0.2}

At $t^\prime=0.2$ a single transition is expected between PM and AFM
phases \cite{Morita02a,Clay08a}.  In Fig.~\ref{tp0.2-result}(a)-(b) we
plot $D$ and $B^\prime$ as a function of $U$. The transition from the
PM to an insulating phase is clearly seen as a discontinuous decrease
in $B^\prime$ and $D$ at $U=U_c$.  $U_c$ is only weakly size dependent
at $t^\prime=0.2$---for the 4$\times$4 lattice\cite{Clay08a} $U_c =
2.95 \pm 0.05$, while for $6\times6$ and $8\times8$ lattices we found
2.75$<U_c<$3.00.
We estimate $U_c$ $\approx$ 2.7 in the thermodynamic limit. 

For all of the $t^\prime$ values we considered, we found that
$d_{x^2-y^2}$ pair-pair correlations were of larger magnitude than
$d_{xy}$ correlations (in Section \ref{results-tp0.8} below we show an
explicit comparison between the two).  Fig.~\ref{tp0.2-result}(c)
shows the $d_{x^2-y^2}$ pair-pair correlations $P_d(r^\star)$ as a
function of $U$.  The distance $r^\star$ is defined as the
next-to-furthest possible separation $r$ between two lattice points on
the finite lattice; $r^\star=2.24$, 3.61, 5.00 for 4$\times$4,
6$\times$6, and 8$\times$8 lattices, respectively.  Here we use
$r^\star$ rather than the furthest distance $r_{\rm max}$ because of
finite-size effects\cite{Aimi07a} associated with $r_{\rm max}$.  The
4$\times$4 correlations are considerably larger in magnitude because
of the larger $r^\star$ on that lattice and we have not included them
on Fig.~\ref{tp0.2-result}(c).  As seen in Fig.~\ref{tp0.2-result}(c),
$P_d(r^\star)$ 
has a tendency to decrease monotonically with $U$  and 
is smaller at all nonzero $U$ compared to $U=0$.
At $U_c$
 $P_d(r^\star)$ 
decreases discontinuously.

In addition to the $U$-dependence, it is also important to examine the
distance-dependence of pair-pair correlations.  In
Fig.~\ref{tp0.2-result}(d) we plot $\Delta P_d(r,U)$, defined as
$\Delta P_d(r,U)=P_d(r,U)-P_d(r,U=0)$, as a function of $r$ for $U$=2.75.
Positive $\Delta P_d(r, U)$ indicates enhanced pairing correlations
over the noninteracting limit. 
 We choose $U=2.75$ in the PM state and
close to the PM-AFM boundary where the greatest enhancement of
pair-pair correlations from AFM fluctuations might be expected from
prior work.  Fig.~\ref{tp0.2-result}(d) includes the exact $4\times4$
$\Delta P_d(r,U)$ as well.  Our results in
Figs.~\ref{tp0.2-result}(c)-(d) show that as the system size
increases, the long-range $d_{x^2-y^2}$ pair-pair correlation function
approaches that of noninteracting fermions.  We have confirmed similar
behavior of $\Delta P_d(r,U)$, viz., absence of enhancement for other
values of $U$ (not shown here) in either the PM or AFM regions.

\begin{figure}
\resizebox{3.0in}{!}{\includegraphics{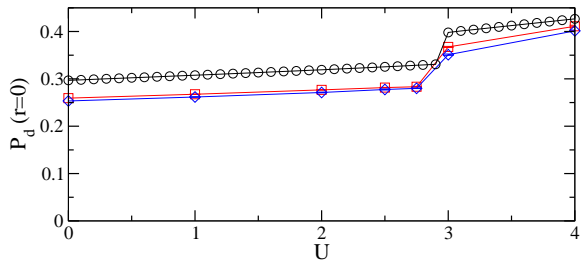}}
\caption{(color online) Short-distance (see text) $d_{x^2-y^2}$
  pair-pair correlation function, $P_d(r=0)$, as a function of $U$ for
  $t^\prime=0.2$.  Circles, squares, and diamonds are for $4\times4$,
  $6\times6$, and $8\times8$ lattices, respectively.  Lines are only guides
  to the eye.}
\label{pr0}
\end{figure}

 As seen in Fig.~\ref{tp0.2-result}(d), the only enhancement by $U$ in
 the pairing correlations is at $r=0$. 
 The $r=0$ enhancement occurs because $P_d(r=0)$ contains a component proportional to the
 nearest-neighbor spin-spin correlation function; the enhancement of
 AFM order by $U$ leads to an increase\cite{Aimi07a} in $P_d(r=0)$.
 In Fig.~\ref{pr0} we plot $P_d(r=0)$ as a function of $U$. Precisely
 at $U_c$ there is a discontinuous {\it increase} in $P_d(r=0)$, even
 as the system becomes {\it semiconducting}, due to the increase in
 the magnitude of AFM spin-spin correlations. 
Importantly, only pair separations of $r>2$ should be used to judge the
enhancement of pairing correlations, because for $r\leq 2$
 $d_{x^2-y^2}$ pairs overlap on the lattice \cite{Clay08a}.
Here we find that 
$P_d(r)$  for distances beyond nearest-neighbor pair separation always {\it
   decrease monotonically with increasing $U$}.
  As we discuss further
 in Section \ref{discussion}, the spurious increase of short-range
 correlations is the primary reason that mean-field calculations find
 SC near the MI transition.

\subsection{$t^\prime=0.5$}
\label{results-tp0.5}

\begin{figure}
\resizebox{3.3in}{!}{\includegraphics{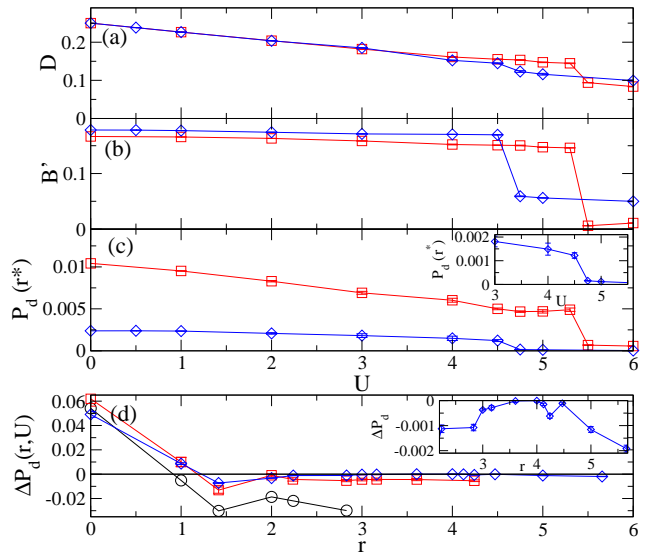}}
\caption{(color online) PIRG results for $t^\prime$=0.5.  Circles,
  squares and diamonds are for 4$\times$4, 6$\times$6 and 8$\times$8
  lattices respectively.  Panels (a)--(c) are the same as
  Fig.~\ref{tp0.2-result}(a)-(c) except $t^\prime$=0.5. 
  (d) is the same as Fig.~\ref{tp0.2-result}(d) except $t^\prime$=0.5 and $U$=4.5.
  In all panels lines are only guides to the eye.}
\label{tp0.5-result}
\end{figure}

 Fig.~\ref{tp0.5-result} shows $D$, $B^\prime$, and $P_d(r)$ for
 $t^\prime=0.5$.  Not surprisingly, compared to $t^\prime=0.2$, $U_{c1}$ 
 depends here more strongly on lattice size, decreasing with
 increasing system size ($U_{c1}=5.4\pm0.1$ and $4.6\pm0.1$ in the
 6$\times$6 and $8\times8$ lattices, respectively).  Previous PIRG
 calculations found $U_{c1}\sim4.1$ after performing finite-size scaling
 \cite{Morita02a}.
 Our results are consistent with this value.

In contradiction to $t^\prime=0.2$ (see Fig.~\ref{tp0.2-result}(b)),
$B^\prime$ here 
is nonzero
on the insulating side of the
MI transition, suggesting that the nature of the insulating phase is
different.  We have also calculated the spin structure factor
$S_\sigma(\vec{q})$ (not shown here). For $U>U_{c1}$, a peak appears in
$S_\sigma(\vec{q})$ at $\vec{q}=(\pi,\pi)$.  However, $S(\pi,\pi)/N$
appears to extrapolate to zero as $N\rightarrow\infty$, based on the
three lattice sizes we have considered. This indicates that the system
does not have long-range AFM order at $t^\prime=0.5$ for $U>U_{c1}$,
consistent with the NMI phase previously identified in this parameter
region \cite{Kashima01a,Morita02a}.  Note that the larger $B^\prime$
in the NMI phase than in the AFM phase is also consistent with our
previous exact diagonalization calculation (see Fig.~2 in
Ref.~\onlinecite{Clay08a}.)  The properties of the NMI phase and the
subsequent NMI-AFM transition at even larger $U_{c2}$ have both been
extensively discussed in the literature before
\cite{Kashima01a,Morita02a,Watanabe03b,Mizusaki06a,Tocchio09a,Yoshioka09a}.
Here therefore we focus on the strength of pair-pair correlations as a
function of $U$ and distance.

Fig.~\ref{tp0.5-result}(c) shows $P_d(r^\star)$ as a function of $U$.
As at $t^\prime=0.2$, $P_d(r^\star)$ decreases monotonically with $U$.
At $U=U_{c1}$, $P_d(r^\star)$ decreases discontinuously and is of very
small magnitude in the NMI phase.  The magnitude of $P_d(r)$ does not
increase as $U$ is increased further approaching the AFM phase.
Fig.~\ref{tp0.5-result}(d) shows $\Delta P_d(r, U)$ as a function of
distance for $t^\prime$=0.5 for $U$=4.5.  As in Fig.~\ref{pr0}, at
$t^\prime=0.5$ $P_d(r=0)$ increases at the MI transition due to the
increase in strength of nearest-neighbor AFM correlations, while
long  distance correlations 
are again weaker at nonzero $U$ than at $U=0$.

\subsection{$t^\prime=0.8$}
\label{results-tp0.8}

\begin{figure}
\resizebox{3.3in}{!}{\includegraphics{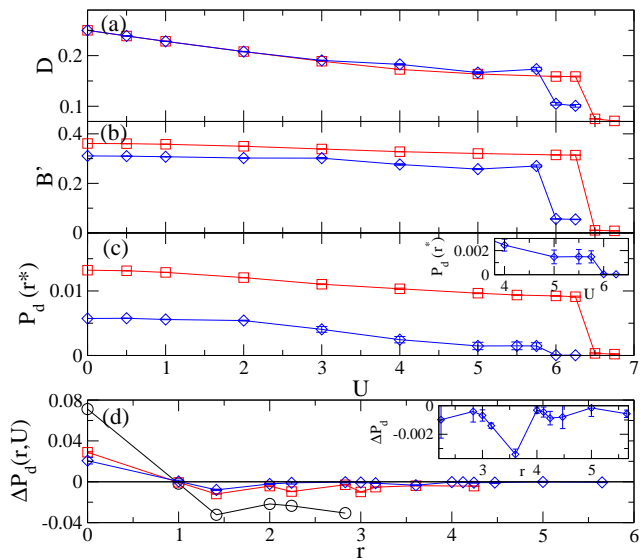}}
\caption{(color online) PIRG results for $t^\prime$=0.8.  Circles,
  squares and diamonds are for 4$\times$4, 6$\times$6 and 8$\times$8
  lattices respectively.  Panels (a)-(c) are the same as
  Fig.~\ref{tp0.5-result}(a)-(c) except $t^\prime$=0.8. 
  In (d) $t^\prime=0.8$ and $U=5.5$. 
  In all panels lines are only guides to the eye.}

\label{tp0.8-result}
\end{figure}

Our results for $t^\prime=0.8$, shown in Fig.~\ref{tp0.8-result}, are
similar to those for $t^\prime=0.5$.  Here the PM region extends to
somewhat larger\cite{Morita02a} $U$.  As at $t^\prime=0.2$ and
$t^\prime=0.5$ there is no enhancement of the pairing correlations.
$P_d(r=0)$ again shows an increase at the MI transition.
 From finite-size scaling we estimated the value for $U_{c1}$ = 5.0 $\pm$ 0.3 from our data. 
 This value is identical to earlier results\cite{Morita02a}.

\begin{figure}[tb]
\resizebox{3.3in}{!}{\includegraphics{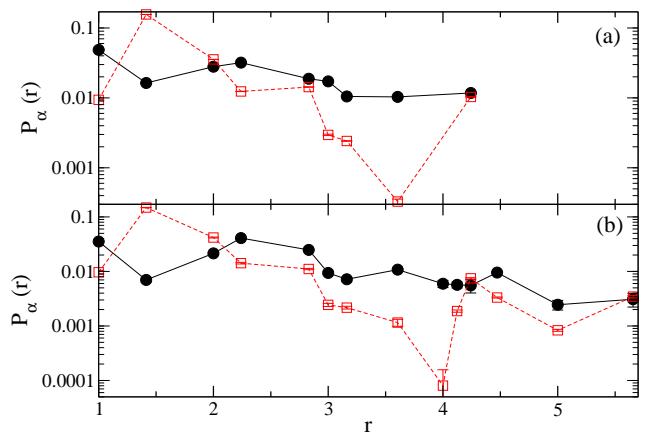}}
\caption{(color online) Comparison of $d_{x^2-y^2}$ and $d_{xy}$
  pair-pair correlations as a function of distance $r$ at
  $t^\prime=0.8$ and $U=4$. (a) 6$\times$6 lattice (b) 8$\times$8
  lattice.  In both panels, circles (squares) correspond to $\alpha=d$
  ($\alpha=d_{xy}$).  Lines are only guides to the eye.}
\label{tp0.8-dxy}
\end{figure}

Several authors have suggested that the symmetry of the
superconducting order parameter changes from $d_{x^2-y^2}$ to $d_{xy}$
or $s+d_{xy}$ in the region of the phase diagram close to the
isotropic triangular lattice limit \cite{Liu05a,Powell07a}. In
Fig.~\ref{tp0.8-dxy} we compare the $d_{x^2-y^2}$ and $d_{xy}$
correlations for $U=4$ as a function of $r$.  Except at specific small
$r$ where pairs can overlap each other on the lattice\cite{Clay08a},
we find that $d_{x^2-y^2}$ correlations are always stronger than
$d_{xy}$ correlations. Plots of the $P_{xy}({\bf r})$ versus $U$ also
show a monotonic decrease with increasing $U$, and $\Delta
P_{xy}(r,U)$ similarly approaches zero for large $r$.

\section{Discussion}
\label{discussion}

In Section \ref{results} we presented superconducting pair-pair
correlations for the ground state of the $\frac{1}{2}$-filled band Hubbard
model on the anisotropic triangular lattice calculated using the PIRG
method.  Our main results are that (i) in all cases the
superconducting pair-pair correlations at all finite $U$ are clearly
weaker than in the noninteracting limit, except for an enhancement of
the very short-range component, and (ii) at large distances the
distance dependence of the pair-pair correlations approaches that of
the noninteracting system. These results, in conjunction with earlier
exact diagonalization results \cite{Clay08a}, which show exactly the
same trends, strongly suggest that the superconductivity is not
present in the model.  Since many of the earlier works did find SC
within the same model Hamiltonian, it is useful to compare these
approaches and results with ours. Broadly speaking, two different
kinds of methods had predicted SC within the triangular lattice
Hubbard model, mean-field and the VMC. We discuss them separately.

\noindent{\it Mean-field approaches:} In all cases, mean-field methods
find a superconducting phase {\it between} the PM and AFM phases
\cite{Kino98a,Kondo98a,Kyung06a,Sahebsara06a}.  A NMI phase is found
by some mean-field methods\cite{Kino98a,Kyung06a,Sahebsara06a} but not
others\cite{Kondo98a}.  However, there are further
inconsistencies---for example, the paramagnetic insulating (PMI) phase
in reference \onlinecite{Kino98a} in some regions of the phase diagram
unrealistically occurs at a temperature {\it higher} than that of the
PM phase.  In the NMI phase the nearest-neighbor AFM correlations are
strong but AFM long-range order is not present.  At fixed $t^\prime$,
increasing $U$ drives the system from the PM to NMI phases.  Upon
entering the NMI the $d_{x^2-y^2}$ correlation at $r=0$ increases in
strength because the $r=0$ correlation is proportional to the
nearest-neighbor spin-spin correlations. Our present results show that
{\it at the same time as this trivial short-range correlation
  increases, the strength of longer-ranged correlations decrease
  greatly}.  Mean-field methods cannot capture these longer-ranged
correlations and erroneously extrapolate from the short-distance
limit.

\noindent{\it VMC:} Within VMC a functional form for the wavefunction
is assumed at the outset.  The three different phases found for
intermediate $t^\prime$ (PM, NMI, and AFM) require three different
assumptions for the functional form of the wavefunction. This makes
determining the true ground state behavior extremely difficult using
VMC, especially near the phase boundaries.  PIRG uses instead an {\it
  unconstrained} superposition of Slater determinants which does not
suffer from these problems.  The assumption of a functional form is a
serious disadvantage as evidenced from the variety of results from
different VMC studies which are not consistent with one another.  For
example, reference \onlinecite{Watanabe06a} did find SC in the model,
while a later study by the same authors did not \cite{Watanabe08a}.
Liu et al. assume that the wavefunction is a Gutzwiller projected BCS
function and find SC \cite{Liu05a}. The appropriateness of such a
wavefunction to describe SC is however a controversial
assumption---Tocchio et al for example did not find SC within the same
assumed wavefunction form \cite{Tocchio09a}.  The occurrence of the NMI
phase within VMC methods seems to be as much of a problem as within
mean-field methods---Tocchio et al do not find evidence for NMI at
$t^\prime$=0.6 while PIRG does \cite{Morita02a,Tocchio09a}.

SC has also been found in several models that are closely related to
the $\frac{1}{2}$-filled Hubbard model on the anisotropic triangular
lattice. These include the Hubbard-Heisenberg model
\cite{Gan05a,Powell05a,Gan06a,Powell07a,Guertler09a,Rau11a}, and the
Hubbard model with {\it two} diagonal $t^\prime$ bonds per square
plaquette \cite{Yokoyama06a,Nevidomskyy08a}. While we cannot compare
directly our PIRG results with these models, in nearly all cases the
methods used to study these models are identical to those that have
erroneously predicted SC within the present model.  We have begun a
reinvestigation of these models.

We now return to the superconducting phase found in the organic CTS.
Our results here cover the entire parameter region ($0.4
\alt t^\prime\alt 0.8$) thought to be
appropriate\cite{Kandpal09a,Nakamura09a} for the $\kappa$-phase CTS
superconductors within the effective Hubbard model description for
them, and clearly indicate that the $\frac{1}{2}$-filled band Hubbard
model is not sufficient to explain the occurrence of SC in
$\kappa$-ET.  It is important in this context to recall that in many
CTS superconductors the insulator-superconductor transition is not
from an AFM phase but from a different kind of exotic
insulator. Examples include $\kappa$-(ET)$_2$Cu$_2$(CN)$_3$, which
lacks long-range AFM order even at the lowest temperatures of
measurement and has been considered a QSL candidate \cite{Kanoda11a},
but is superconducting under pressure \cite{Kurosaki05a}, and other
CTS superconductors where the insulating phase adjacent to
superconductivity is nonmagnetic and charge-ordered (CO)
\cite{Dressel11a}.  Once again mean-field theory, now within the
$\frac{1}{4}$-field band extended Hubbard Hamiltonian on anisotropic
triangular lattices has suggested the possibility of a charge
fluctuation mediated CO--SC transition \cite{Merino01a}.  Based on our
present work, there are reasons to doubt mean-field approaches to SC
within correlated-electron models in general and these earlier results
should be checked through many-body calculations.

The different natures of the unconventional semiconductors proximate
to SC in the CTS confront theorists with a unique challenge. While
mean-field theories suggest a different mechanism for each different
semiconductor-superconductor transition, it appears unlikely to us
that structurally similar materials, with identical molecular
components in some cases, should require different mechanisms for SC.
Given how difficult a problem correlated-electron SC has turned out to
be we suggest that an alternate approach is to determine first how to
construct a theoretical framework within which a unified theory of SC
begins to look feasible, and then to search for the same.  We have
recently shown how it may be possible to construct such a framework for the CTS
\cite{Li10a,Dayal11a,Mazumdar11a,Mazumdar12a}.  In this picture, the
$\kappa$-(ET)$_2$X and other dimerized CTS should be described in
terms of the {\it underlying $\frac{1}{4}$-filled band} as with the
other CTS superconductors \cite{Li10a,Dayal11a}. In the presence of
strong dimerization and relatively weak frustration, AFM wins. Under
increasing frustration though, a transition occurs from AFM with
uniform charge density on each dimer to a charge-ordered paired
singlet state we have termed the Paired Electron Crystal (PEC)
\cite{Li10a,Dayal11a}.  Experimental examples of the PEC in 2D CTS
include $\beta$-({\it meso}-DMBEDT-TTF)$_2$PF$_6$ (reference
\onlinecite{Kimura06a}) and $\beta^\prime$-EtMe$_3$P[Pd(dmit)$_2$]$_2$
(references \onlinecite{Tamura09a,Yamamoto11a}), {\it which have
  precisely the same CO and bond patterns as in the PEC model
  \cite{Li10a,Dayal11a},} and are superconducting under pressure.  The
application of pressure corresponds to a further increase in
frustration and gives the possibility of a paired electron {\it
  liquid} superconductor \cite{Mazumdar08a}, a realization of the
charged boson SC first proposed by Schafroth \cite{Schafroth55a}.
Although more work will be necessary to prove this,
this theoretical approach has the advantage that it leads to a single
model for correlated-electron SC in the CTS. Even more interestingly,
we have pointed out that there exist several frustrated strongly
correlated {\it inorganic} $\frac{1}{4}$-filled superconductors that
can perhaps be described within the same model
\cite{Dayal11a,Mazumdar11a,Mazumdar12a}.

Finally, the experimental observation of AFM \cite{Iwasa03a} in
expanded fullerides A$_3$C$_{60}$ has led to the modeling of these
compounds in terms of a 3D nondegenerate $\frac{1}{2}$-filled band
Hubbard model \cite{Capone09a}. The threefold degeneracy of the lowest
antibonding molecular orbitals in C$_{60}$ is removed by Jahn-Teller
instability \cite{Iwasa03a,Capone09a}.  The observation of a spingap
in the antiferromagnetic state validates the nondegenerate description
\cite{Capone09a}. The dynamic mean-field theory (DMFT) proposed for
the AFM to SC transition in the fullerides within this 3D effective
$\frac{1}{2}$-filled band Hubbard model \cite{Capone09a} is however
very similar to the DMFT theories of SC in the 2D CTS
\cite{Kyung06a}. Our results here suggest that a reexamination of the
spin-fluctuation mechanism of SC in the fullerides may also be called
for.

\section{Acknowledgments}

This work was supported by the US Department of Energy grant
DE-FG02-06ER46315. RTC thanks the Institute for Solid State Physics of
the University of Tokyo for hospitality while on sabbatical where
a portion of this work was completed.


\begin{thebibliography}{61}
\expandafter\ifx\csname natexlab\endcsname\relax\def\natexlab#1{#1}\fi
\expandafter\ifx\csname bibnamefont\endcsname\relax
  \def\bibnamefont#1{#1}\fi
\expandafter\ifx\csname bibfnamefont\endcsname\relax
  \def\bibfnamefont#1{#1}\fi
\expandafter\ifx\csname citenamefont\endcsname\relax
  \def\citenamefont#1{#1}\fi
\expandafter\ifx\csname url\endcsname\relax
  \def\url#1{\texttt{#1}}\fi
\expandafter\ifx\csname urlprefix\endcsname\relax\def\urlprefix{URL }\fi
\providecommand{\bibinfo}[2]{#2}
\providecommand{\eprint}[2][]{\url{#2}}

\bibitem[{\citenamefont{Kanoda and Kato}(2011)}]{Kanoda11a}
\bibinfo{author}{\bibfnamefont{K.}~\bibnamefont{Kanoda}} \bibnamefont{and}
  \bibinfo{author}{\bibfnamefont{R.}~\bibnamefont{Kato}},
  \bibinfo{journal}{Annu. Rev. Condens. Matter Phys.}
  \textbf{\bibinfo{volume}{2}}, \bibinfo{pages}{167} (\bibinfo{year}{2011}).

\bibitem[{\citenamefont{Kashima and Imada}(2001{\natexlab{a}})}]{Kashima01a}
\bibinfo{author}{\bibfnamefont{T.}~\bibnamefont{Kashima}} \bibnamefont{and}
  \bibinfo{author}{\bibfnamefont{M.}~\bibnamefont{Imada}},
  \bibinfo{journal}{J.\ Phys.\ Soc.\ Jpn.} \textbf{\bibinfo{volume}{70}},
  \bibinfo{pages}{3052} (\bibinfo{year}{2001}{\natexlab{a}}).

\bibitem[{\citenamefont{Morita et~al.}(2002)\citenamefont{Morita, Watanabe, and
  Imada}}]{Morita02a}
\bibinfo{author}{\bibfnamefont{H.}~\bibnamefont{Morita}},
  \bibinfo{author}{\bibfnamefont{S.}~\bibnamefont{Watanabe}}, \bibnamefont{and}
  \bibinfo{author}{\bibfnamefont{M.}~\bibnamefont{Imada}},
  \bibinfo{journal}{J.\ Phys.\ Soc.\ Jpn.} \textbf{\bibinfo{volume}{71}},
  \bibinfo{pages}{2109} (\bibinfo{year}{2002}).

\bibitem[{\citenamefont{Watanabe}(2003)}]{Watanabe03b}
\bibinfo{author}{\bibfnamefont{S.}~\bibnamefont{Watanabe}},
  \bibinfo{journal}{J.\ Phys.\ Soc.\ Jpn.} \textbf{\bibinfo{volume}{72}},
  \bibinfo{pages}{2042} (\bibinfo{year}{2003}).

\bibitem[{\citenamefont{Parcollet et~al.}(2004)\citenamefont{Parcollet, Biroli,
  and Kotliar}}]{Parcollet04a}
\bibinfo{author}{\bibfnamefont{O.}~\bibnamefont{Parcollet}},
  \bibinfo{author}{\bibfnamefont{G.}~\bibnamefont{Biroli}}, \bibnamefont{and}
  \bibinfo{author}{\bibfnamefont{G.}~\bibnamefont{Kotliar}},
  \bibinfo{journal}{Phys.\ Rev.\ Lett.} \textbf{\bibinfo{volume}{92}},
  \bibinfo{pages}{226402} (\bibinfo{year}{2004}).

\bibitem[{\citenamefont{Mizusaki and Imada}(2006)}]{Mizusaki06a}
\bibinfo{author}{\bibfnamefont{T.}~\bibnamefont{Mizusaki}} \bibnamefont{and}
  \bibinfo{author}{\bibfnamefont{M.}~\bibnamefont{Imada}},
  \bibinfo{journal}{Phys.\ Rev.\ B} \textbf{\bibinfo{volume}{74}},
  \bibinfo{eid}{014421} (\bibinfo{year}{2006}).

\bibitem[{\citenamefont{Kyung and Tremblay}(2006)}]{Kyung06a}
\bibinfo{author}{\bibfnamefont{B.}~\bibnamefont{Kyung}} \bibnamefont{and}
  \bibinfo{author}{\bibfnamefont{A.~M.~S.} \bibnamefont{Tremblay}},
  \bibinfo{journal}{Phys.\ Rev.\ Lett.} \textbf{\bibinfo{volume}{97}},
  \bibinfo{pages}{046402} (\bibinfo{year}{2006}).

\bibitem[{\citenamefont{Sahebsara and
  \protect{S\'en\'echal}}(2006)}]{Sahebsara06a}
\bibinfo{author}{\bibfnamefont{P.}~\bibnamefont{Sahebsara}} \bibnamefont{and}
  \bibinfo{author}{\bibfnamefont{D.}~\bibnamefont{\protect{S\'en\'echal}}},
  \bibinfo{journal}{Phys.\ Rev.\ Lett.} \textbf{\bibinfo{volume}{97}},
  \bibinfo{pages}{257004} (\bibinfo{year}{2006}).

\bibitem[{\citenamefont{Koretsune et~al.}(2007)\citenamefont{Koretsune, Motome,
  and Furusaki}}]{Koretsune07a}
\bibinfo{author}{\bibfnamefont{T.}~\bibnamefont{Koretsune}},
  \bibinfo{author}{\bibfnamefont{Y.}~\bibnamefont{Motome}}, \bibnamefont{and}
  \bibinfo{author}{\bibfnamefont{A.}~\bibnamefont{Furusaki}},
  \bibinfo{journal}{J.\ Phys.\ Soc.\ Jpn.} \textbf{\bibinfo{volume}{76}},
  \bibinfo{pages}{074719} (\bibinfo{year}{2007}).

\bibitem[{\citenamefont{Clay et~al.}(2008)\citenamefont{Clay, Li, and
  Mazumdar}}]{Clay08a}
\bibinfo{author}{\bibfnamefont{R.~T.} \bibnamefont{Clay}},
  \bibinfo{author}{\bibfnamefont{H.}~\bibnamefont{Li}}, \bibnamefont{and}
  \bibinfo{author}{\bibfnamefont{S.}~\bibnamefont{Mazumdar}},
  \bibinfo{journal}{Phys.\ Rev.\ Lett.} \textbf{\bibinfo{volume}{101}},
  \bibinfo{pages}{166403} (\bibinfo{year}{2008}).

\bibitem[{\citenamefont{Lee et~al.}(2008)\citenamefont{Lee, Li, and
  Monien}}]{Lee08b}
\bibinfo{author}{\bibfnamefont{H.}~\bibnamefont{Lee}},
  \bibinfo{author}{\bibfnamefont{G.}~\bibnamefont{Li}}, \bibnamefont{and}
  \bibinfo{author}{\bibfnamefont{H.}~\bibnamefont{Monien}},
  \bibinfo{journal}{Phys.\ Rev.\ B} \textbf{\bibinfo{volume}{78}},
  \bibinfo{pages}{205117} (\bibinfo{year}{2008}).

\bibitem[{\citenamefont{Sahebsara and
  \protect{S\'en\'echal}}(2008)}]{Sahebsara08a}
\bibinfo{author}{\bibfnamefont{P.}~\bibnamefont{Sahebsara}} \bibnamefont{and}
  \bibinfo{author}{\bibfnamefont{D.}~\bibnamefont{\protect{S\'en\'echal}}},
  \bibinfo{journal}{Phys.\ Rev.\ Lett.} \textbf{\bibinfo{volume}{100}},
  \bibinfo{pages}{136402} (\bibinfo{year}{2008}).

\bibitem[{\citenamefont{Watanabe et~al.}(2008)\citenamefont{Watanabe, Yokoyama,
  Tanaka, and Inoue}}]{Watanabe08a}
\bibinfo{author}{\bibfnamefont{T.}~\bibnamefont{Watanabe}},
  \bibinfo{author}{\bibfnamefont{H.}~\bibnamefont{Yokoyama}},
  \bibinfo{author}{\bibfnamefont{Y.}~\bibnamefont{Tanaka}}, \bibnamefont{and}
  \bibinfo{author}{\bibfnamefont{J.}~\bibnamefont{Inoue}},
  \bibinfo{journal}{Phys.\ Rev.\ B} \textbf{\bibinfo{volume}{77}},
  \bibinfo{pages}{214505} (\bibinfo{year}{2008}).

\bibitem[{\citenamefont{Ohashi et~al.}(2008)\citenamefont{Ohashi, Momoi,
  Tsunetsugu, and Kawakami}}]{Ohashi08a}
\bibinfo{author}{\bibfnamefont{T.}~\bibnamefont{Ohashi}},
  \bibinfo{author}{\bibfnamefont{T.}~\bibnamefont{Momoi}},
  \bibinfo{author}{\bibfnamefont{H.}~\bibnamefont{Tsunetsugu}},
  \bibnamefont{and} \bibinfo{author}{\bibfnamefont{N.}~\bibnamefont{Kawakami}},
  \bibinfo{journal}{Phys.\ Rev.\ Lett.} \textbf{\bibinfo{volume}{100}},
  \bibinfo{pages}{076402} (\bibinfo{year}{2008}).

\bibitem[{\citenamefont{Tocchio et~al.}(2009)\citenamefont{Tocchio, Parola,
  Gros, and Becca}}]{Tocchio09a}
\bibinfo{author}{\bibfnamefont{L.~F.} \bibnamefont{Tocchio}},
  \bibinfo{author}{\bibfnamefont{A.}~\bibnamefont{Parola}},
  \bibinfo{author}{\bibfnamefont{C.}~\bibnamefont{Gros}}, \bibnamefont{and}
  \bibinfo{author}{\bibfnamefont{F.}~\bibnamefont{Becca}},
  \bibinfo{journal}{Phys.\ Rev.\ B} \textbf{\bibinfo{volume}{80}},
  \bibinfo{pages}{064419} (\bibinfo{year}{2009}).

\bibitem[{\citenamefont{Yoshioka et~al.}(2009)\citenamefont{Yoshioka, Koga, and
  Kawakami}}]{Yoshioka09a}
\bibinfo{author}{\bibfnamefont{T.}~\bibnamefont{Yoshioka}},
  \bibinfo{author}{\bibfnamefont{A.}~\bibnamefont{Koga}}, \bibnamefont{and}
  \bibinfo{author}{\bibfnamefont{N.}~\bibnamefont{Kawakami}},
  \bibinfo{journal}{Phys.\ Rev.\ Lett.} \textbf{\bibinfo{volume}{103}},
  \bibinfo{pages}{036401} (\bibinfo{year}{2009}).

\bibitem[{\citenamefont{Guertler et~al.}(2009)\citenamefont{Guertler, Wang, and
  Zhang}}]{Guertler09a}
\bibinfo{author}{\bibfnamefont{S.}~\bibnamefont{Guertler}},
  \bibinfo{author}{\bibfnamefont{Q.~H.} \bibnamefont{Wang}}, \bibnamefont{and}
  \bibinfo{author}{\bibfnamefont{F.~C.} \bibnamefont{Zhang}},
  \bibinfo{journal}{Phys.\ Rev.\ B} \textbf{\bibinfo{volume}{79}},
  \bibinfo{pages}{144526} (\bibinfo{year}{2009}).

\bibitem[{\citenamefont{Liebsch et~al.}(2009)\citenamefont{Liebsch, Ishida, and
  Merino}}]{Liebsch09a}
\bibinfo{author}{\bibfnamefont{A.}~\bibnamefont{Liebsch}},
  \bibinfo{author}{\bibfnamefont{H.}~\bibnamefont{Ishida}}, \bibnamefont{and}
  \bibinfo{author}{\bibfnamefont{J.}~\bibnamefont{Merino}},
  \bibinfo{journal}{Phys.\ Rev.\ B} \textbf{\bibinfo{volume}{79}},
  \bibinfo{pages}{195108} (\bibinfo{year}{2009}).

\bibitem[{\citenamefont{Yu and Yin}(2010)}]{Yu10a}
\bibinfo{author}{\bibfnamefont{Z.~Q.} \bibnamefont{Yu}} \bibnamefont{and}
  \bibinfo{author}{\bibfnamefont{L.}~\bibnamefont{Yin}},
  \bibinfo{journal}{Phys.\ Rev.\ B} \textbf{\bibinfo{volume}{81}},
  \bibinfo{pages}{195122} (\bibinfo{year}{2010}).

\bibitem[{\citenamefont{Schmalian}(1998)}]{Schmalian98a}
\bibinfo{author}{\bibfnamefont{J.}~\bibnamefont{Schmalian}},
  \bibinfo{journal}{Phys.\ Rev.\ Lett.} \textbf{\bibinfo{volume}{81}},
  \bibinfo{pages}{4232} (\bibinfo{year}{1998}).

\bibitem[{\citenamefont{Kino and Kontani}(1998)}]{Kino98a}
\bibinfo{author}{\bibfnamefont{H.}~\bibnamefont{Kino}} \bibnamefont{and}
  \bibinfo{author}{\bibfnamefont{H.}~\bibnamefont{Kontani}},
  \bibinfo{journal}{J.\ Phys.\ Soc.\ Jpn.} \textbf{\bibinfo{volume}{67}},
  \bibinfo{pages}{3691} (\bibinfo{year}{1998}).

\bibitem[{\citenamefont{Kondo and Moriya}(1998)}]{Kondo98a}
\bibinfo{author}{\bibfnamefont{H.}~\bibnamefont{Kondo}} \bibnamefont{and}
  \bibinfo{author}{\bibfnamefont{T.}~\bibnamefont{Moriya}},
  \bibinfo{journal}{J.\ Phys.\ Soc.\ Jpn.} \textbf{\bibinfo{volume}{67}},
  \bibinfo{pages}{3695} (\bibinfo{year}{1998}).

\bibitem[{\citenamefont{Vojta and Dagotto}(1999)}]{Vojta99a}
\bibinfo{author}{\bibfnamefont{M.}~\bibnamefont{Vojta}} \bibnamefont{and}
  \bibinfo{author}{\bibfnamefont{E.}~\bibnamefont{Dagotto}},
  \bibinfo{journal}{Phys.\ Rev.\ B} \textbf{\bibinfo{volume}{59}},
  \bibinfo{pages}{R713} (\bibinfo{year}{1999}).

\bibitem[{\citenamefont{Baskaran}(2003)}]{Baskaran03a}
\bibinfo{author}{\bibfnamefont{G.}~\bibnamefont{Baskaran}},
  \bibinfo{journal}{Phys.\ Rev.\ Lett.} \textbf{\bibinfo{volume}{90}},
  \bibinfo{pages}{197007} (\bibinfo{year}{2003}).

\bibitem[{\citenamefont{Liu et~al.}(2005)\citenamefont{Liu, Schmalian, and
  Trivedi}}]{Liu05a}
\bibinfo{author}{\bibfnamefont{J.}~\bibnamefont{Liu}},
  \bibinfo{author}{\bibfnamefont{J.}~\bibnamefont{Schmalian}},
  \bibnamefont{and} \bibinfo{author}{\bibfnamefont{N.}~\bibnamefont{Trivedi}},
  \bibinfo{journal}{Phys.\ Rev.\ Lett.} \textbf{\bibinfo{volume}{94}},
  \bibinfo{pages}{127003} (\bibinfo{year}{2005}).

\bibitem[{\citenamefont{Nevidomskyy et~al.}(2008)\citenamefont{Nevidomskyy,
  Scheiber, S\'en\'echal, and Tremblay}}]{Nevidomskyy08a}
\bibinfo{author}{\bibfnamefont{A.~H.} \bibnamefont{Nevidomskyy}},
  \bibinfo{author}{\bibfnamefont{C.}~\bibnamefont{Scheiber}},
  \bibinfo{author}{\bibfnamefont{D.}~\bibnamefont{S\'en\'echal}},
  \bibnamefont{and} \bibinfo{author}{\bibfnamefont{A.-M.~S.}
  \bibnamefont{Tremblay}}, \bibinfo{journal}{Phys.\ Rev.\ B}
  \textbf{\bibinfo{volume}{77}}, \bibinfo{pages}{064427}
  (\bibinfo{year}{2008}).

\bibitem[{\citenamefont{Gan et~al.}(2005)\citenamefont{Gan, Chen, Su, and
  Zhang}}]{Gan05a}
\bibinfo{author}{\bibfnamefont{J.~Y.} \bibnamefont{Gan}},
  \bibinfo{author}{\bibfnamefont{Y.}~\bibnamefont{Chen}},
  \bibinfo{author}{\bibfnamefont{Z.~B.} \bibnamefont{Su}}, \bibnamefont{and}
  \bibinfo{author}{\bibfnamefont{F.~C.} \bibnamefont{Zhang}},
  \bibinfo{journal}{Phys.\ Rev.\ Lett.} \textbf{\bibinfo{volume}{94}},
  \bibinfo{pages}{067005} (\bibinfo{year}{2005}).

\bibitem[{\citenamefont{Powell and McKenzie}(2005)}]{Powell05a}
\bibinfo{author}{\bibfnamefont{B.~J.} \bibnamefont{Powell}} \bibnamefont{and}
  \bibinfo{author}{\bibfnamefont{R.~H.} \bibnamefont{McKenzie}},
  \bibinfo{journal}{Phys.\ Rev.\ Lett.} \textbf{\bibinfo{volume}{94}},
  \bibinfo{pages}{047004} (\bibinfo{year}{2005}).

\bibitem[{\citenamefont{Gan et~al.}(2006)\citenamefont{Gan, Chen, and
  Zhang}}]{Gan06a}
\bibinfo{author}{\bibfnamefont{J.~Y.} \bibnamefont{Gan}},
  \bibinfo{author}{\bibfnamefont{Y.}~\bibnamefont{Chen}}, \bibnamefont{and}
  \bibinfo{author}{\bibfnamefont{F.~C.} \bibnamefont{Zhang}},
  \bibinfo{journal}{Phys.\ Rev.\ B} \textbf{\bibinfo{volume}{74}},
  \bibinfo{pages}{094515} (\bibinfo{year}{2006}).

\bibitem[{\citenamefont{Powell and McKenzie}(2007)}]{Powell07a}
\bibinfo{author}{\bibfnamefont{B.~J.} \bibnamefont{Powell}} \bibnamefont{and}
  \bibinfo{author}{\bibfnamefont{R.~H.} \bibnamefont{McKenzie}},
  \bibinfo{journal}{Phys.\ Rev.\ Lett.} \textbf{\bibinfo{volume}{98}},
  \bibinfo{pages}{027005} (\bibinfo{year}{2007}).

\bibitem[{\citenamefont{Rau and Kee}(2011)}]{Rau11a}
\bibinfo{author}{\bibfnamefont{J.~G.} \bibnamefont{Rau}} \bibnamefont{and}
  \bibinfo{author}{\bibfnamefont{H.-Y.} \bibnamefont{Kee}},
  \bibinfo{journal}{Phys.\ Rev.\ Lett.} \textbf{\bibinfo{volume}{106}},
  \bibinfo{pages}{056405} (\bibinfo{year}{2011}).

\bibitem[{\citenamefont{Kandpal et~al.}(2009)\citenamefont{Kandpal, Opahle,
  Zhang, Jeschke, and Valent\'\i}}]{Kandpal09a}
\bibinfo{author}{\bibfnamefont{H.~C.} \bibnamefont{Kandpal}},
  \bibinfo{author}{\bibfnamefont{I.}~\bibnamefont{Opahle}},
  \bibinfo{author}{\bibfnamefont{Y.-Z.} \bibnamefont{Zhang}},
  \bibinfo{author}{\bibfnamefont{H.~O.} \bibnamefont{Jeschke}},
  \bibnamefont{and}
  \bibinfo{author}{\bibfnamefont{R.}~\bibnamefont{Valent\'\i}},
  \bibinfo{journal}{Phys.\ Rev.\ Lett.} \textbf{\bibinfo{volume}{103}},
  \bibinfo{pages}{067004} (\bibinfo{year}{2009}).

\bibitem[{\citenamefont{Nakamura et~al.}(2009)\citenamefont{Nakamura,
  Yoshimoto, Kosugi, Arita, and Imada}}]{Nakamura09a}
\bibinfo{author}{\bibfnamefont{K.}~\bibnamefont{Nakamura}},
  \bibinfo{author}{\bibfnamefont{Y.}~\bibnamefont{Yoshimoto}},
  \bibinfo{author}{\bibfnamefont{T.}~\bibnamefont{Kosugi}},
  \bibinfo{author}{\bibfnamefont{R.}~\bibnamefont{Arita}}, \bibnamefont{and}
  \bibinfo{author}{\bibfnamefont{M.}~\bibnamefont{Imada}},
  \bibinfo{journal}{J.\ Phys.\ Soc.\ Jpn.} \textbf{\bibinfo{volume}{78}},
  \bibinfo{pages}{083710} (\bibinfo{year}{2009}).

\bibitem[{\citenamefont{Watanabe et~al.}(2006)\citenamefont{Watanabe, Yokoyama,
  Tanaka, and Inoue}}]{Watanabe06a}
\bibinfo{author}{\bibfnamefont{T.}~\bibnamefont{Watanabe}},
  \bibinfo{author}{\bibfnamefont{H.}~\bibnamefont{Yokoyama}},
  \bibinfo{author}{\bibfnamefont{Y.}~\bibnamefont{Tanaka}}, \bibnamefont{and}
  \bibinfo{author}{\bibfnamefont{J.}~\bibnamefont{Inoue}},
  \bibinfo{journal}{J.\ Phys.\ Soc.\ Jpn.} \textbf{\bibinfo{volume}{75}},
  \bibinfo{pages}{074707} (\bibinfo{year}{2006}).

\bibitem[{\citenamefont{Yokoyama et~al.}(2006)\citenamefont{Yokoyama, Ogata,
  and Tanaka}}]{Yokoyama06a}
\bibinfo{author}{\bibfnamefont{H.}~\bibnamefont{Yokoyama}},
  \bibinfo{author}{\bibfnamefont{M.}~\bibnamefont{Ogata}}, \bibnamefont{and}
  \bibinfo{author}{\bibfnamefont{Y.}~\bibnamefont{Tanaka}},
  \bibinfo{journal}{J.\ Phys.\ Soc.\ Jpn.} \textbf{\bibinfo{volume}{75}},
  \bibinfo{pages}{114706} (\bibinfo{year}{2006}).

\bibitem[{\citenamefont{Imada and Kashima}(2000)}]{Imada00a}
\bibinfo{author}{\bibfnamefont{M.}~\bibnamefont{Imada}} \bibnamefont{and}
  \bibinfo{author}{\bibfnamefont{T.}~\bibnamefont{Kashima}},
  \bibinfo{journal}{J.\ Phys.\ Soc.\ Jpn.} \textbf{\bibinfo{volume}{69}},
  \bibinfo{pages}{2723} (\bibinfo{year}{2000}).

\bibitem[{\citenamefont{Kashima and Imada}(2001{\natexlab{b}})}]{Kashima01b}
\bibinfo{author}{\bibfnamefont{T.}~\bibnamefont{Kashima}} \bibnamefont{and}
  \bibinfo{author}{\bibfnamefont{M.}~\bibnamefont{Imada}},
  \bibinfo{journal}{J.\ Phys.\ Soc.\ Jpn.} \textbf{\bibinfo{volume}{70}},
  \bibinfo{pages}{2287} (\bibinfo{year}{2001}{\natexlab{b}}).

\bibitem[{\citenamefont{Mizusaki and Imada}(2004)}]{Mizusaki04a}
\bibinfo{author}{\bibfnamefont{T.}~\bibnamefont{Mizusaki}} \bibnamefont{and}
  \bibinfo{author}{\bibfnamefont{M.}~\bibnamefont{Imada}},
  \bibinfo{journal}{Phys.\ Rev.\ B} \textbf{\bibinfo{volume}{69}},
  \bibinfo{pages}{125110} (\bibinfo{year}{2004}).

\bibitem[{\citenamefont{Imada and Mizusaki}(2006)}]{Imada06a}
\bibinfo{author}{\bibfnamefont{M.}~\bibnamefont{Imada}} \bibnamefont{and}
  \bibinfo{author}{\bibfnamefont{T.}~\bibnamefont{Mizusaki}}, in
  \emph{\bibinfo{booktitle}{Effective models for low-dimensional
  strongly-correlated systems}}, edited by
  \bibinfo{editor}{\bibfnamefont{G.~G.} \bibnamefont{Batrouni}}
  \bibnamefont{and} \bibinfo{editor}{\bibfnamefont{D.}~\bibnamefont{Poilblanc}}
  (\bibinfo{publisher}{AIP Conference Proceedings}, \bibinfo{address}{New
  York}, \bibinfo{year}{2006}), vol. \bibinfo{volume}{816}, pp.
  \bibinfo{pages}{78--91}.

\bibitem[{\citenamefont{Miyagawa et~al.}(1995)\citenamefont{Miyagawa, Kawamoto,
  Nakazawa, and Kanoda}}]{Miyagawa95a}
\bibinfo{author}{\bibfnamefont{K.}~\bibnamefont{Miyagawa}},
  \bibinfo{author}{\bibfnamefont{A.}~\bibnamefont{Kawamoto}},
  \bibinfo{author}{\bibfnamefont{Y.}~\bibnamefont{Nakazawa}}, \bibnamefont{and}
  \bibinfo{author}{\bibfnamefont{K.}~\bibnamefont{Kanoda}},
  \bibinfo{journal}{Phys.\ Rev.\ Lett.} \textbf{\bibinfo{volume}{75}},
  \bibinfo{pages}{1174} (\bibinfo{year}{1995}).

\bibitem[{\citenamefont{Miyagawa et~al.}(2002)\citenamefont{Miyagawa, Kawamoto,
  and Kanoda}}]{Miyagawa02a}
\bibinfo{author}{\bibfnamefont{K.}~\bibnamefont{Miyagawa}},
  \bibinfo{author}{\bibfnamefont{A.}~\bibnamefont{Kawamoto}}, \bibnamefont{and}
  \bibinfo{author}{\bibfnamefont{K.}~\bibnamefont{Kanoda}},
  \bibinfo{journal}{Phys.\ Rev.\ Lett.} \textbf{\bibinfo{volume}{89}},
  \bibinfo{pages}{017003} (\bibinfo{year}{2002}).

\bibitem[{\citenamefont{Scalettar et~al.}(1989)\citenamefont{Scalettar, Loh,
  Gubernatis, Moreo, White, Scalapino, Sugar, and Dagotto}}]{Scalettar89b}
\bibinfo{author}{\bibfnamefont{R.~T.} \bibnamefont{Scalettar}},
  \bibinfo{author}{\bibfnamefont{E.~Y.} \bibnamefont{Loh}},
  \bibinfo{author}{\bibfnamefont{J.~E.} \bibnamefont{Gubernatis}},
  \bibinfo{author}{\bibfnamefont{A.}~\bibnamefont{Moreo}},
  \bibinfo{author}{\bibfnamefont{S.~R.} \bibnamefont{White}},
  \bibinfo{author}{\bibfnamefont{D.~J.} \bibnamefont{Scalapino}},
  \bibinfo{author}{\bibfnamefont{R.~L.} \bibnamefont{Sugar}}, \bibnamefont{and}
  \bibinfo{author}{\bibfnamefont{E.}~\bibnamefont{Dagotto}},
  \bibinfo{journal}{Phys.\ Rev.\ Lett.} \textbf{\bibinfo{volume}{62}},
  \bibinfo{pages}{1407} (\bibinfo{year}{1989}).

\bibitem[{\citenamefont{Moreo et~al.}(1992)\citenamefont{Moreo, Scalapino, and
  White}}]{Moreo92b}
\bibinfo{author}{\bibfnamefont{A.}~\bibnamefont{Moreo}},
  \bibinfo{author}{\bibfnamefont{D.~J.} \bibnamefont{Scalapino}},
  \bibnamefont{and} \bibinfo{author}{\bibfnamefont{S.~R.} \bibnamefont{White}},
  \bibinfo{journal}{Phys.\ Rev.\ B} \textbf{\bibinfo{volume}{45}},
  \bibinfo{pages}{7544} (\bibinfo{year}{1992}).

\bibitem[{\citenamefont{Huscroft and Scalettar}(1998)}]{Huscroft98a}
\bibinfo{author}{\bibfnamefont{C.}~\bibnamefont{Huscroft}} \bibnamefont{and}
  \bibinfo{author}{\bibfnamefont{R.~T.} \bibnamefont{Scalettar}},
  \bibinfo{journal}{Phys.\ Rev.\ Lett.} \textbf{\bibinfo{volume}{81}},
  \bibinfo{pages}{2775} (\bibinfo{year}{1998}).

\bibitem[{\citenamefont{Aimi and Imada}(2007)}]{Aimi07a}
\bibinfo{author}{\bibfnamefont{T.}~\bibnamefont{Aimi}} \bibnamefont{and}
  \bibinfo{author}{\bibfnamefont{M.}~\bibnamefont{Imada}},
  \bibinfo{journal}{J.\ Phys.\ Soc.\ Jpn.} \textbf{\bibinfo{volume}{76}},
  \bibinfo{pages}{113708} (\bibinfo{year}{2007}).

\bibitem[{\citenamefont{Yoshioka
  et~al.}(2008{\natexlab{a}})\citenamefont{Yoshioka, Koga, and
  Kawakami}}]{Yoshioka08a}
\bibinfo{author}{\bibfnamefont{T.}~\bibnamefont{Yoshioka}},
  \bibinfo{author}{\bibfnamefont{A.}~\bibnamefont{Koga}}, \bibnamefont{and}
  \bibinfo{author}{\bibfnamefont{N.}~\bibnamefont{Kawakami}},
  \bibinfo{journal}{J.\ Phys.\ Soc.\ Jpn.} \textbf{\bibinfo{volume}{77}},
  \bibinfo{pages}{104702} (\bibinfo{year}{2008}{\natexlab{a}}).

\bibitem[{\citenamefont{Yoshioka
  et~al.}(2008{\natexlab{b}})\citenamefont{Yoshioka, Koga, and
  Kawakami}}]{Yoshioka08b}
\bibinfo{author}{\bibfnamefont{T.}~\bibnamefont{Yoshioka}},
  \bibinfo{author}{\bibfnamefont{A.}~\bibnamefont{Koga}}, \bibnamefont{and}
  \bibinfo{author}{\bibfnamefont{N.}~\bibnamefont{Kawakami}},
  \bibinfo{journal}{Phys.\ Rev.\ B} \textbf{\bibinfo{volume}{78}},
  \bibinfo{pages}{165113} (\bibinfo{year}{2008}{\natexlab{b}}).

\bibitem[{\citenamefont{Kurosaki et~al.}(2005)\citenamefont{Kurosaki, Shimizu,
  Miyagawa, Kanoda, and Saito}}]{Kurosaki05a}
\bibinfo{author}{\bibfnamefont{Y.}~\bibnamefont{Kurosaki}},
  \bibinfo{author}{\bibfnamefont{Y.}~\bibnamefont{Shimizu}},
  \bibinfo{author}{\bibfnamefont{K.}~\bibnamefont{Miyagawa}},
  \bibinfo{author}{\bibfnamefont{K.}~\bibnamefont{Kanoda}}, \bibnamefont{and}
  \bibinfo{author}{\bibfnamefont{G.}~\bibnamefont{Saito}},
  \bibinfo{journal}{Phys. Rev. Lett.} \textbf{\bibinfo{volume}{95}},
  \bibinfo{pages}{177001} (\bibinfo{year}{2005}).

\bibitem[{\citenamefont{Dressel}(2011)}]{Dressel11a}
\bibinfo{author}{\bibfnamefont{M.}~\bibnamefont{Dressel}}, \bibinfo{journal}{J.
  Phys.: Condens. Matter} \textbf{\bibinfo{volume}{23}},
  \bibinfo{pages}{293201} (\bibinfo{year}{2011}).

\bibitem[{\citenamefont{Merino and McKenzie}(2001)}]{Merino01a}
\bibinfo{author}{\bibfnamefont{J.}~\bibnamefont{Merino}} \bibnamefont{and}
  \bibinfo{author}{\bibfnamefont{R.~H.} \bibnamefont{McKenzie}},
  \bibinfo{journal}{Phys.\ Rev.\ Lett.} \textbf{\bibinfo{volume}{87}},
  \bibinfo{pages}{237002} (\bibinfo{year}{2001}).

\bibitem[{\citenamefont{Li et~al.}(2010)\citenamefont{Li, Clay, and
  Mazumdar}}]{Li10a}
\bibinfo{author}{\bibfnamefont{H.}~\bibnamefont{Li}},
  \bibinfo{author}{\bibfnamefont{R.~T.} \bibnamefont{Clay}}, \bibnamefont{and}
  \bibinfo{author}{\bibfnamefont{S.}~\bibnamefont{Mazumdar}},
  \bibinfo{journal}{J. Phys.: Condens. Matter} \textbf{\bibinfo{volume}{22}},
  \bibinfo{pages}{272201} (\bibinfo{year}{2010}).

\bibitem[{\citenamefont{Dayal et~al.}(2011)\citenamefont{Dayal, Clay, Li, and
  Mazumdar}}]{Dayal11a}
\bibinfo{author}{\bibfnamefont{S.}~\bibnamefont{Dayal}},
  \bibinfo{author}{\bibfnamefont{R.~T.} \bibnamefont{Clay}},
  \bibinfo{author}{\bibfnamefont{H.}~\bibnamefont{Li}}, \bibnamefont{and}
  \bibinfo{author}{\bibfnamefont{S.}~\bibnamefont{Mazumdar}},
  \bibinfo{journal}{Phys.\ Rev.\ B} \textbf{\bibinfo{volume}{83}},
  \bibinfo{pages}{245106} (\bibinfo{year}{2011}).

\bibitem[{\citenamefont{Mazumdar and Clay}(2011)}]{Mazumdar11a}
\bibinfo{author}{\bibfnamefont{S.}~\bibnamefont{Mazumdar}} \bibnamefont{and}
  \bibinfo{author}{\bibfnamefont{R.~T.} \bibnamefont{Clay}}
  \bibinfo{journal}{Phys. Status Solidi B} (\bibinfo{year}{2012}), \bibinfo{note}{in press, 
  corrected proof http://dx.doi.org/10.1002/pssb.201100723}.

\bibitem[{\citenamefont{Mazumdar et~al.}(2012)\citenamefont{Mazumdar, Clay, and
  Li}}]{Mazumdar12a}
\bibinfo{author}{\bibfnamefont{S.}~\bibnamefont{Mazumdar}},
  \bibinfo{author}{\bibfnamefont{R.~T.} \bibnamefont{Clay}}, \bibnamefont{and}
  \bibinfo{author}{\bibfnamefont{H.}~\bibnamefont{Li}},
  \bibinfo{journal}{Physica B}  (\bibinfo{year}{2012}), \bibinfo{note}{in
  press, corrected proof http://dx.doi.org/10.1016/j.physb.2012.01.016}.

\bibitem[{\citenamefont{Kimura et~al.}(2006)\citenamefont{Kimura, Suzuki,
  Maejima, Mori, Yamaura, Kakiuchi, Sawa, and Moriyama}}]{Kimura06a}
\bibinfo{author}{\bibfnamefont{S.}~\bibnamefont{Kimura}},
  \bibinfo{author}{\bibfnamefont{H.}~\bibnamefont{Suzuki}},
  \bibinfo{author}{\bibfnamefont{T.}~\bibnamefont{Maejima}},
  \bibinfo{author}{\bibfnamefont{H.}~\bibnamefont{Mori}},
  \bibinfo{author}{\bibfnamefont{J.}~\bibnamefont{Yamaura}},
  \bibinfo{author}{\bibfnamefont{T.}~\bibnamefont{Kakiuchi}},
  \bibinfo{author}{\bibfnamefont{H.}~\bibnamefont{Sawa}}, \bibnamefont{and}
  \bibinfo{author}{\bibfnamefont{H.}~\bibnamefont{Moriyama}},
  \bibinfo{journal}{J. Am. Chem. Soc.} \textbf{\bibinfo{volume}{128}},
  \bibinfo{pages}{1456} (\bibinfo{year}{2006}).

\bibitem[{\citenamefont{Tamura and Kato}(2009)}]{Tamura09a}
\bibinfo{author}{\bibfnamefont{M.}~\bibnamefont{Tamura}} \bibnamefont{and}
  \bibinfo{author}{\bibfnamefont{R.}~\bibnamefont{Kato}},
  \bibinfo{journal}{Sci. Technol. Adv. Mater.} \textbf{\bibinfo{volume}{10}},
  \bibinfo{pages}{024304} (\bibinfo{year}{2009}).

\bibitem[{\citenamefont{Yamamoto et~al.}(2011)\citenamefont{Yamamoto, Nakazawa,
  Tamura, Nakao, Ikemoto, Moriwaki, Fukaya, Kato, and Yakushi}}]{Yamamoto11a}
\bibinfo{author}{\bibfnamefont{T.}~\bibnamefont{Yamamoto}},
  \bibinfo{author}{\bibfnamefont{Y.}~\bibnamefont{Nakazawa}},
  \bibinfo{author}{\bibfnamefont{M.}~\bibnamefont{Tamura}},
  \bibinfo{author}{\bibfnamefont{A.}~\bibnamefont{Nakao}},
  \bibinfo{author}{\bibfnamefont{Y.}~\bibnamefont{Ikemoto}},
  \bibinfo{author}{\bibfnamefont{T.}~\bibnamefont{Moriwaki}},
  \bibinfo{author}{\bibfnamefont{A.}~\bibnamefont{Fukaya}},
  \bibinfo{author}{\bibfnamefont{R.}~\bibnamefont{Kato}}, \bibnamefont{and}
  \bibinfo{author}{\bibfnamefont{K.}~\bibnamefont{Yakushi}},
  \bibinfo{journal}{J.\ Phys.\ Soc.\ Jpn.} \textbf{\bibinfo{volume}{80}},
  \bibinfo{pages}{123709} (\bibinfo{year}{2011}).

\bibitem[{\citenamefont{Mazumdar and Clay}(2008)}]{Mazumdar08a}
\bibinfo{author}{\bibfnamefont{S.}~\bibnamefont{Mazumdar}} \bibnamefont{and}
  \bibinfo{author}{\bibfnamefont{R.~T.} \bibnamefont{Clay}},
  \bibinfo{journal}{Phys.\ Rev.\ B} \textbf{\bibinfo{volume}{77}},
  \bibinfo{pages}{180515(R)} (\bibinfo{year}{2008}).

\bibitem[{\citenamefont{Schafroth}(1955)}]{Schafroth55a}
\bibinfo{author}{\bibfnamefont{M.~R.} \bibnamefont{Schafroth}},
  \bibinfo{journal}{Phys. Rev.} \textbf{\bibinfo{volume}{100}},
  \bibinfo{pages}{463} (\bibinfo{year}{1955}).

\bibitem[{\citenamefont{Iwasa and Takenobu}(2003)}]{Iwasa03a}
\bibinfo{author}{\bibfnamefont{Y.}~\bibnamefont{Iwasa}} \bibnamefont{and}
  \bibinfo{author}{\bibfnamefont{T.}~\bibnamefont{Takenobu}},
  \bibinfo{journal}{J. Phys.: Condens. Matter} \textbf{\bibinfo{volume}{15}},
  \bibinfo{pages}{R495} (\bibinfo{year}{2003}).

\bibitem[{\citenamefont{Capone et~al.}(2009)\citenamefont{Capone, Fabrizio,
  Castellani, and Tosatti}}]{Capone09a}
\bibinfo{author}{\bibfnamefont{M.}~\bibnamefont{Capone}},
  \bibinfo{author}{\bibfnamefont{M.}~\bibnamefont{Fabrizio}},
  \bibinfo{author}{\bibfnamefont{C.}~\bibnamefont{Castellani}},
  \bibnamefont{and} \bibinfo{author}{\bibfnamefont{E.}~\bibnamefont{Tosatti}},
  \bibinfo{journal}{Rev.\ Mod.\ Phys.} \textbf{\bibinfo{volume}{81}},
  \bibinfo{pages}{943} (\bibinfo{year}{2009}).

\end{thebibliography}

\end{document}